\documentclass[prc,reprint,a4paper,twoside,showpacs,amsmath]
  {revtex4-1}

\usepackage[colorlinks=true,allcolors=blue]{hyperref}

\newcommand\etal{\textit{~et~al.}}
\newcommand\je{j_\text{e}}
\newcommand\jeb{\bar\jmath_\text{e}}
\newcommand\jd{j_\text{d}}
\newcommand\jdb{\bar\jmath_\text{d}}
\newcommand\Sp{\mathit{Sp}}
\newcommand\bra\langle
\newcommand\ket\rangle
\newcommand\ds\displaystyle
\newcommand\qud{\hspace{5pt}}
\newcommand\qudd{\qud\phantom{0}}
\newcommand\quddd{\qud\phantom{00}}

\begin{document}

\title{Symplectic group structure of the $^{48}$Cr, $^{88}$Ru, and
  $^{92}$Pd ground states}

\author{K. Neerg\aa rd}{
\affiliation{Fjordtoften 17, 4700 N\ae stved, Denmark}

\begin{abstract}
  The ground states of $^{48}$Cr, $^{88}$Ru, and $^{92}$Pd are studied
  in the $1f_{7/2}$ or $1g_{9/2}$ shell model with effective
  interactions from the literature. They are found to be composed,
  quite independently of the shell and the interaction, roughly of
  75\% of $(s,t)=(0,0)$ and 25\% of $(s,t)=(4,0)$, where $s$ is the
  seniority and $t$ the reduced isospin. Other irreps of the
  symplectic group $\Sp(2j+1)$, where $j$ is the single-nucleon
  angular momentum, make only very small contributions. The state
  $\chi$ obtained by antisymmetrization and normalization of the
  ground state in the stretch scheme of Danos and Gillet [M. Danos and
  V. Gillet, Phys. Rev. 161, 1034 (1967)] has a very different
  structure where the $\Sp(2j+1)$ irreps other than $(s,t)=(0,0)$ and
  $(4,0)$ contribute 20\% and 41\% for $j=7/2$ and $9/2$,
  respectively. The contributions of $\chi$ and the $s=0$ state to the
  calculated states are about equal for $^{48}$Cr. For $^{88}$Ru and
  $^{92}$Pd the $s=0$ state is unambigously a better approximation to
  the calculated states than $\chi$. A state $\chi'$ obtained by
  antisymmetrization and normalization of the product of two
  stretch-scheme ground states of the system with two valence nucleons
  or nucleon holes of each type has much larger overlaps with the
  calculated ground states than $\chi$ but a deviating $\Sp(2j+1)$
  decomposition.
\end{abstract}


\maketitle

\section{\label{sc:intr}Introduction}

In an impressive experiment, Cederwall\etal~\cite{Ce11} measured the
gamma decay of three excited states of $^{92}$Pd, which has four
neutrons and four protons less than the doubly magic $^{100}$Sn. They
interpreted the spectrum in terms of the ``stretch scheme'' proposed
in the 1960s by Danos and Gillet~\cite{DaGi67} to describe deformed
nuclei in the shell model. Qi\etal~\cite{Qi11} made shell model
calculations in support of this interpretation employing a valence
space composed of the shells $2p_{3/2}$, $1f_{5/2}$, $2p_{1/2}$, and
$1g_{7/2}$ as well as smaller spaces}. The stretch scheme applies to
nuclei with equal even numbers of valence neutrons and protons. All
valence neutron orbits are supposed to belong to the same $j$ shell
and so also for the protons. The valence nucleons are divided into two
``chains,'' each of them formed by half of the valence neutrons and
half of the valence protons. Within a chain the nucleonic angular
momenta are coupled to the maximal total angular momentum through
pairwise coupling of a neutron and a proton to their maximal combined
angular momentum. In the ground state the chain angular momenta are
opposite, and ``rotational'' excitations are formed by a bending of
them towards each other to form a nonzero total angular momentum.
Nucleon holes in a $j$ shell may replace valence nucleons without
changing the scheme essentially. This is its version relevant to
$^{92}$Pd, which may be seen as a system of four neutron holes and
four proton holes in the $1g_{9/2}$ shell. In the following I use the
term \textit{quasinucleon} to denote either a valence nucleon or a
nucleon hole, and I call two quasinucleons \textit{equivalent} if
either both of them are valence nucleons or both of them are nucleon
holes.

Generalizing the adaption to nuclei independently by Bohr, Mottelson
and Pines~\cite{BoMoPi58} and Bogolyubov and Solov'yov~\cite{BoSO58}
of the theory of superconductivity of Bardeen, Cooper, and
Schrieffer~\cite{BaCoSchr57}, Goswami and Kisslinger~\cite{GoKi65}
introduced in the 1960s a concept of ``isoscalar pairing'' different
from the ``isovector pairing'' described by the
Bardeen-Cooper-Schrieffer theory. This concept is much discussed in
the subsequent literature; see the review by Frauendorf and
Macchiavelli~\cite{FrMa14}. Referring to predictions of isoscalar
pairing in nuclei with equal numbers $N$ and $Z$ of neutrons end
protons, Cederwall\etal~\cite{Ce11} state that their results ``reveal
evidence for a spin-aligned, isoscalar neutronproton coupling scheme''
and ``suggest that this coupling scheme replaces normal superfluidity
(characterized by seniority coupling) in the ground and lowest excited
states of the heaviest $N=Z$ nuclei.''

In support of this suggestion, Qi\etal~\cite{Qi11} point out that in
single-$j$-shell calculations for the system of two quasineutrons and
two equivalent quasiprotons in the $1f_{7/2}$, $1g_{9/2}$, or
$1h_{11/2}$ shell with empirical effective interaction, the product of
the state of two quasineutrons with combined angular momentum zero and
the similar state of the two quasiprotons makes up only a little more
than half of the calculated ground states. Consideration of this
product is motivated by its resemblance to the product of neutron and
proton Bardeen-Cooper-Schrieffer states conventionally employed to
model nuclear superfluidity. It is, however, not an eigenstate of
isospin. In a meaningful adaption of the concept of nuclear
superfluidity to the single-$j$-shell model of $N=Z$ nuclei one should
rather see the unique state with isospin and seniority zero as the
manifestation of isovector pairing. I show in a previous
article~\cite{Ne13} that this state makes up about 80\% of the
calculated ground states of two quasineutrons and two equivalent
quasiprotons in the $1f_{7/2}$ or $1g_{9/2}$ shell.

The stretch-scheme ground state is not antisymmetric in the
quasineutrons or the quasiprotons. Qi\etal~\cite{Qi11} consider the
antisymmetrized and normalized state and get overlaps of 92--95\% with
the calculated ground states of two quasineutrons and two equivalent
quasiprotons. This finding is essentially confirmed by my calculations
in Ref.~\cite{Ne13}. I find, moreover, that the overlaps are larger in
the $1f_{7/2}$ shell than in the $1g_{9/2}$ shell. The overlap of the
seniority zero state with the antisymmetized and normalized
stretch-scheme state is 62\% and 52\%, respectively, so to this extent
seniority zero and the stretch scheme are different visualizations of
the same physics, in this case of two quasineutrons and two equivalent
quasiprotons.

In their analysis of calculated states of $^{92}$Pd,
Qi\etal~\cite{Qi11} counted the numbers of pairs of $1g_{9/2}$ holes
with definite combined angular momentum. I show in Ref.~\cite{Ne13}
that for two quasiprotons and two quasineutrons in the $1g_{9/2}$
shell, the seniority zero state has a large content of quasinucleon
pairs with high angular momenta, and the antisymmetrized and
normalized stretch-scheme state has a large content of pairs with low
angular momenta. Inferring a pairing mode from such counts is thus not
straightforward.

The present text presents an analysis similar to the one in
Ref.~\cite{Ne13} addressing the case of $^{92}$Pd. I thus consider the
system of four quasineutrons and four equivalent quasiprotons in a
single $j$ shell. Besides $^{92}$Pd this is the single-$j$-shell
configuration of its $1g_{9/2}$ cross conjugate, $^{88}$Ru, and of
$^{48}$Cr in the $1f_{7/2}$ shell. The particle-hole symmetry of the
single-$j$-shell model is, in the $1f_{7/2}$ shell, only approximately
obeyed by the data. Van Isacker~\cite{Is13}, in his $1f_{7/2}$ shell
model calculations, accordingly makes an interpolation between the
empirical two-valence-nucleon and two-nucleon-hole interactions. As
seen from Table~\ref{tb:tb},
\begin{table*}
  \caption{\label{tb:tb}Expectation values and overlaps in
    percentages. Due to rounding off the sum of these percentages may
    differ slightly from 100 and a zero only means that the
    percentage is less than 0.5. The operator $P_{s,t}$ is the
    projection onto the subspace of the $I=T=0$ space with the given
    $(s,t)$ and $\bra P_{s,t}\ket_a$ is its expectation value in state
    $a$. State $\psi$ is the calculated ground state and states
    $\chi$ and $\chi'$ are defined by Eqs.~\eqref{eq:chi}
    and~\eqref{eq:chi'}. In the lasts four columns pairs of
    percentages are shown. The first percentage is for $a=\chi$ and
    the second one is for $a=\chi'$. The rows ``$\psi=\chi$'' and
    ``$\psi=\chi'$'' show the $\Sp(2j+1)$ decompositions of these
    states and the row ``Dimension'' shows the subspace dimension.}
\begin{ruledtabular}
\begin{tabular}{lrrrrrrrrr}
&$\bra P_{0,0}\ket_\psi$&$\bra P_{4,0}\ket_\psi$
&$\bra P_{4,2}\ket_\psi$&$\bra P_{6,1}\ket_\psi$
&$\bra P_{8,0}\ket_\psi$
&\hspace{3em}$|\bra a|\psi\ket|^2$
&$\ds\frac{|\bra a|P_{4,0}|\psi\ket|^2}
  {\bra P_{4,0}\ket_a\bra P_{4,0}\ket_\psi}$
&$\ds\frac{|\bra a|P_{6,1}|\psi\ket|^2}
  {\bra P_{6,1}\ket_a\bra P_{6,1}\ket_\psi}$
&$\ds\frac{|\bra a|P_{8,0}|\psi\ket|^2}
  {\bra P_{8,0}\ket_a\bra P_{8,0}\ket_\psi}$\\[-10pt]
\\
\hline\\[-8pt]
$^{48}$Cr\\[3pt]
Dimension&1&2&0&2&1\\[5pt]
SchTr, emp.&75&24&&0&1&78\qudd 93&97\qudd 98\\
SchTr, fit&79&20&&0&0&77\qudd 93&97\qudd 98\\
ZR I&80&20&&0&0&77\qudd 92&98\qudd 98\\
ZR II&73&26&&0&1&80\qudd 92&100\qudd 94\\[3pt]
$\psi=\chi$&49&30&&19&2\\
$\psi=\chi'$&67&26&&6&1\\[12pt]
($^{88}$Ru,) $^{92}$Pd\\[3pt]
Dimension&1&2&1&5&7\\[3pt]
SchTr, emp.&70&27&1&0&2&65\qudd 86&97\qudd 99&77\qudd 36&86\qudd 98\\
SchTr, fit&70&27&0&0&2&63\qudd 87&95\qud 100&76\qudd 36&85\qudd 98\\
QLW&70&27&1&0&2&66\qudd 84&99\qudd 96&71\qudd 39&89\qudd 96\\
ZE I&83&16&0&0&1&50\qudd 80&93\qud 100&1\qudd 93&82\qudd 99\\
ZE II&72&25&0&0&2&61\qudd 86&97\qudd 99&61\qudd 46&86\qudd 98\\
ZE III&85&14&1&0&1&44\qudd 76&100\qudd 93&36\quddd 7&89\qudd 95\\
ZE IV&76&22&0&0&1&61\qudd 82&99\qudd 97&60\qudd 46&88\qudd 97\\
CCGI&76&22&0&0&2&56\qudd 84&92\qud 100&49\qudd 63&82\qudd 98\\
SLGT0&73&25&0&0&2&63\qudd 85&96\qudd 99&75\qudd 40&86\qudd 98\\
GF&68&28&1&0&3&67\qudd 86&97\qudd 99&83\qudd 31&87\qudd 98\\
Nb90&70&27&1&0&2&64\qudd 86&96\qudd 99&77\qudd 36&86\qudd 98\\[3pt]
$\psi=\chi$&25&34&10&20&10\\
$\psi=\chi'$&47&36&0&6&10
\end{tabular}
\end{ruledtabular}
\end{table*}
these interactions, denoted there by ZR~I and II, give qualitatively
similar results in the present type of analysis. Effects breaking the
particle-hole symmetry are thus apparently minorly important in this
context. The situation is somewhat different in the $1g_{9/2}$ shell
because the observed $^{80}$Zr spectrum is clearly rotational and thus
not that of a closed-shell nucleus. The nucleus $^{88}$Ru seems to the
sit on the edge of an onset of deformation with $N=Z$ decreasing from
50, so modeling it by the $1g_{9/2}$ shell model may be questionable.
The focus of my study as concerns the $1g_{9/2}$ shell is on $^{92}$Pd.

In the next Sec.~\ref{sc:meth} I describe the method used to construct
the interaction matrix in the space of isospin and angular momentum
zero and decompose the calculated ground state into irreps of the
symplectic group $\Sp(2j+1)$, where $j$ is the single-nucleon angular
momentum. The results of this decomposition are shown and discussed in
Sec.~\ref{sc:dec}. Section~\ref{sc:str} discusses the stretch-scheme
ground state. It is found that only a small part of it belongs to the
space of states antisymmetric in the quasineutrons and in the
quasiprotons. Following Qi\etal~\cite{Qi11} I antisymmetrize and
normalize this part and then discuss the decomposition of the
antisymmetrized and normalized state into irreps of $\Sp(2j+1)$ and
its overlaps with the calculated states. In Sec.~\ref{sc:oth} a
similar analysis is applied to the state obtained by
antisymmetrization and normalization of the product of two
stretch-scheme ground states of the system of two quasineutrons and
two equivalent quasiprotons. The article is summarized in
Sec.~\ref{sc:sum}.

\section{\label{sc:meth}Method}

The eight quasinucleons are labeled with numbers 1--8 so that
quasinucleons 1--4 are quasineutrons and quasinucleons 5--8 are
quasiprotons. The angular momentum of the $i$th quasinucleon is
denoted by $j_i$, and all these angular momenta are equal to $j=7/2$
in the $1f_{7/2}$ shell and $j=9/2$ in the $1g_{9/2}$ shell. States of
the system with total angular momentum $I=0$ may be expanded on a
basis of states:
\begin{equation}\label{eq:ba}
  |\alpha\ket=|[(\je\beta)_n(\je\gamma)_p]0\ket\,,
\end{equation}
where $|\je\beta\ket_n$ is a totally antisymmetric state with angular
momentum $\je$ of the quasineutrons. The index $\beta$ labels a
complete, orthonormal set of such states. The definition of
$|\je\gamma\ket_p$ is analogous for quasiprotons. The outmost (square)
brackets in Eq.~\eqref{eq:ba} followed by the value of $I$ indicate
vector coupling with the total magnetic quantum number suppressed. A
similar notation is employed thoughout this article.

The states $|\je\beta\ket_n$ may be expanded on a basis of states:
\begin{equation}\label{eq:bae}
  |j_{12}j_{34}\ket_{\je}=|[(j_1j_2)j_{12}(j_3j_4)j_{34}]\je\ket\,,
\end{equation}
with even $j_{12}$ and $j_{34}$. To determine the subspaces with a
given symmetry of the span of this basis one can use that the sum of
transpositions
\begin{equation}\label{eq:K4}
  K_4=\sum_{1\le i<k\le4}(ik)
\end{equation}
is in the symmetric group $S(4)$, a class sum, and therefore within
each irrep a constant dependent only on the irrep. Because the
states~\eqref{eq:bae} carry the irrep $[1^2]\times[1^2]$ of the
product of the $S(2)$ of quasinucleons 1 and 2 and quasinucleons 3 and
4, the Young frames of the irreps of $S(4)$ present in their span have
at most two columns. Let such a frame have column lengths $\lambda$
and $\mu$. One can calculate its $K_4$ by evaluating in
Yamanouchi's~\cite{Ya37} realization of the irrep the diagonal matrix
element of the sum~\eqref{eq:K4} in the tableau where the indices 1--4
appear successively from top to bottom in the columns from left to
right. The result, which I denote by just $K$ because it is not
limited to the case $n=\lambda+\mu=4$, is
\begin{equation}\label{eq:K}
  K=n-n^2/4-d(d+1)\,,
\end{equation}
with $d=(\lambda-\mu)/2$. Because $K$ as given by this expression
decreases with $d$, different $(\lambda,\mu)$ with the same $n$ have
different $K$. Indicating by a prime the restriction of operators to
the span of the states~\eqref{eq:bae} with a given $\je$, we have
\begin{equation}\label{eq:K4'}
  K_4'=-2+4(23)'\,.
\end{equation}
The subspaces of definite symmetry are thus the eigen\-spaces of
$(23)'$. The totally antisymmetric states $|\je\beta\ket_n$, which
have $(\lambda,\mu)=(4,0)$, in particular form a basis for the
eigenspace with eigenvalue $-1$. They are therefore obtained by
diagonalization of $(23)'$ in the basis~\eqref{eq:bae}. The matrix
elements are
\begin{equation}\label{(23)'}
  \bra j_{12}j_{34}|(23)'|j_{12}'j_{34}'\ket_{\je}
  =\bra j_{12}j_{34}|j_{12}'j_{34}'\ket_{jjjj\je}
\end{equation}
in terms of what Zamick and Escuderos~\cite{ZaEs13} call the unitary
nine-$j$ symbol:
\begin{equation}\label{9j}
  \bra ef|gh \ket_{abcdi}
  =\bra[(ab)e(cd)f]i|[(ac)g(bd)h]i\ket\,.
\end{equation}

A charge-independent interaction of two quasinucleons in the same $j$
shell can be written
\begin{equation}\label{eq:V}
  V=\sum_J c_J P_J
\end{equation}
with
\begin{equation}\label{eq:PJ}
  P_J=\sum_{1\le i<k\le8} P_{j_{ik}=J}\,,
\end{equation}
where $P_{j_{ik}=J}$ denotes the projection onto the eigenspace with
eigenvalue $J$ of the combined angular momentum $j_{ik}$ of the $i$th
and $k$th quasinucleons. Indicating by a double prime the restriction
of operators to the span of the states~\eqref{eq:ba}, we have
\begin{equation}\label{eq:PJ'}
  P_J''=12P_{j_{12}=J}''+16P_{j_{15}=J}''\,.
\end{equation}
In the basis of states
\begin{equation}\label{eq:baj}
  |j_{12}j_{34}j_{56}j_{78}\je\ket
  =|[(j_{12}j_{34})_{\je}(j_{56}j_{78})_{\je}]0\ket
\end{equation}
the projection $P_{j_{12}=J}$ is diagonal with matrix elements
$\delta_{j_{12}J}$. The projection $P_{j_{15}=J}$ has the matrix
elements
\begin{multline}\label{eq:P15me}
  \bra j_{12}j_{34}j_{56}j_{78}\je|P_{j_{15}=J}
    |j_{12}'j_{34}'j_{56}'j_{78}'\je'\ket\\
  =\delta_{j_{34}j_{34}'}\delta_{j_{78}j_{78}'}\\
    \sum_{j_{26},\jd}\bra j_{12}j_{56}\je|Jj_{26}\jd\ket_{j_{34}j_{78}}
      \bra j_{12}'j_{56}'\je'|Jj_{26}\jd\ket_{j_{34}j_{78}}\,,
\end{multline}
with
\begin{multline}\label{eq:x}
  \bra j_{12}j_{56}\je|Jj_{26}\jd\ket_{j_{34}j_{78}}\\
  =\bra\je\je|\jd\jd\ket_{j_{12}j_{34}j_{56}j_{78}0}
    \bra j_{12}j_{56}|Jj_{26}\ket_{jjjj\jd}\,.
\end{multline}

Because the states~\eqref{eq:ba} carry the irrep $[1^4]\times[1^4]$ of
the product of the quasineutron and the quasiproton $S(4)$, the Young
frames of the irreps of $S(8)$ present in their span have at most two
columns. In Eq.~\eqref{eq:K} we now have $n=8$, and $d$ is the isospin
$T$. By the relation
\begin{equation}\label{eq:K8''}
  K_8''=-12+16(15)''\,,
\end{equation}
where
\begin{equation}\label{eq:K8}
  K_8=\sum_{1\le i<k\le8}(ik)\,,
\end{equation}
the subspaces with definite $T$ are thus obtained by diagonalization
of $(15)''$. In particular the $T=0$ space has $(15)''=1/4$. The
matrix elements of $(15)''$ are obtained from
\begin{equation}\label{eq:(15)''}
  (15)''=-\sum_J(-)^JP_{j_{15}=J}''
\end{equation}
and the restriction of Eq.~\eqref{eq:P15me} to the span of the
states~\eqref{eq:ba}.

Each eigenspace of $T$ is the intersection with the $I=0$ space of an
irrep of the unitary group $U(2j+1)$ characterized by $n$ and $T$, and
these irreps split into irreps of $\Sp(2j+1)$ characterized by a
seniority $s$ and a reduced isospin $t$~\cite{Fl52}. Racah's seniority
operator~\cite{Ra43}, generalized to $jj$ coupling and nuclei by
Edmonds and Flowers~\cite{EdFl52},
\begin{equation}\label{eq:Q}
  Q=(2j+1)P_0
\end{equation}
is within each such irrep a constant dependent for a given $j$ only on
the $U(2j+1)$ and $\Sp(2j+1)$ irreps. Edmonds and
Flowers~\cite{EdFl52} derive a closed expression which can be written
\begin{equation}\label{eq:Qx}
  Q=f(j,n,T)-f(j,s,t)\,,
\end{equation}
with
\begin{equation}\label{eq:f}
  f(j,x,y)=(j+2)x-x^2/4-y(y+1)\,.
\end{equation}
Using Flowers's method~\cite{Fl52} one finds that $(n,T)=(8,0)$ is
composed for $j\ge7/2$ of $(s,t)=(0,0)$, $(2,1)$, $(4,0)$, $(4,2)$,
$(6,1)$, and $(8,0)$. These are seen from Eqs.~\eqref{eq:Qx} and
\eqref{eq:f} to have distinct $Q$. The corresponding subspaces of the
$I=T=0$ space are therefore obtained by diagonalization of the
restriction of the operator~\eqref{eq:Q}. Because $(s,t)=(2,1)$ is
composed of $I=2,4,\dots,2j-1$~\cite{Fl52}, its intersection with the
$I=T=0$ space is the null space.

\section{\label{sc:dec}$\Sp(2j+1)$ decompositions of calculated ground
  states}

Calculations were made with the same effective $1f_{7/2}$ and
$1g_{9/2}$ interactions as in Ref.~\cite{Ne13}. Thence I repeat a
brief description of each of them. The interactions SchTr are from the
appendix of the classic study by Schiffer and True~\cite{SchTr76} with
``emp.'' referring to the empirical matrix elements and ``fit'' to those
derived from a universal interaction fitted to the data. ZR~I and II
are Models~I and II of Zamick and Robinson~\cite{ZaRo00}. They were
derived from the spectra of $^{42}$Sc and~$^{54}$Co, respectively. QLW
is $0g_{9/2}$ of Qi, Liotta, and Wyss~\cite{Qi12}. It was extracted
from an interaction for the $2p_{1/2}+1g_{9/2}$ configuration space
provided by Johnstone and Skouras~\cite{JoSk01}. ZE~I--IV are from
Zamick and Escuderos~\cite{ZaEs12}. Specifically, ZE~I and~II are their
INTc and INTd. The former consists of a $T=1$ part from the spectrum
of $^{98}$Cd and a $T=0$ part from a delta interaction. The latter has
a lower $c_9$. ZE~III and~IV are from the spectrum of $^{90}$Nb with
different choices of the $1^+$ level. CCGI is adapted from the
$V_{\text{low-}k}$ of Coraggio, Covello, Gargano, and
Itaco~\cite{CoCoGaIt12}. This is not charge independent. To conserve
isospin, I use their neutron-proton matrix elements in all channels.
SLGT0, GF, and Nb90 (named Nb90ZI in Ref.~\cite{Ne13}) are from
Zerguine and~Van Isacker~\cite{ZeIs11}. Specifically, SLGT0 and GF
were constructed by renormalization to the $1g_{9/2}$ subspace of
interactions for the $2p_{1/2}+1g_{9/2}$ configuration space provided,
respectively, by Serduke, Lawson, and Gloeckner~\cite{Se76}, and Gross
and Frenkel~\cite{GrFr76}, and Nb90 is from the spectrum of $^{90}$Nb
with yet another choice of $1^+$ level.

Table~\ref{tb:tb} shows for each interaction the decomposition of the
ground state into $\Sp(2j+1)$ irreps. It is seen that quite
independently of the shell and the interaction the ground state is
composed roughly of 75\% of $(s,t)=(0,0)$ and 25\% of $(s,t)=(4,0)$.
Other irreps make only small contributions, which tend, however, to be
somewhat larger in the $1g_{9/2}$ shell than in the $1f_{7/2}$ shell.
The typical contribution of about 75\% of the $s=0$ state found here
for $n=8$ is slightly less than the typical 80\% found in
Ref.~\cite{Ne13} for $n=4$. Yet this state, which, as explained in the
Introduction, may be conceived of as the manifestation of perfect
isovector pairing in the single-$j$-shell model of $N=Z$ nuclei,
remains a fairly good first approximation also for $n=8$.

Comparison with the case $n=4$ studied in Ref.~\cite{Ne13} reveals a
striking similarity: In that case only $(s,t)=(0,0)$ and $(4,0)$
occur; for $n=8$ they dominate the calculated states in about the same
ratio. A hint to an understanding of this similarity may conceivably
be found in Qi's calculations~\cite{Qi12a} with the interaction QWL of
multihole states in the $1g_{9/2}$ shell, which show that the
antisymmetrized and normalized product of a pair of $^{96}$Gd ground
states makes up in this model 96\% of the $^{92}$Pd ground state.
(Clearly from comparison with Ref.~\cite{Qi11} the quantity denoted by
$x^2$ in Table~I of Ref.~\cite{Qi12a} is just $x$.)

Because a two-quasinucleon interaction can break at most two $J=0$
pairs, its matrix elements between $\Sp(2j+1)$ irreps differing by
more than four in $s$ vanish. In an expansion where the terms in the
interaction~\eqref{eq:V} other than the pairing force, $J=0$, are
treated as perturbations, the ground state components with $s>4$ are
therefore of second order. This explains their small size. That the
$(s,t)=(4,2)$ component in the $1g_{9/2}$ shells are much smaller than
the $(s,t)=(4,0)$ components is due to smaller matrix elements from
$s=0$. For $j=7/2$ all matrix elements involving $(s,t)=(6,1)$, and
therefore this component, vanish within the numeric accuracy for all
the interactions. Some fundamental selection rule thus seems to be
active in this case. The same is not true for $j=9/2$ and I have no
explanation for this apparent partial conservation of seniority, which
bears a resemblance to the much discussed case of
$j=9/2,n=2T=s=2t=4$, and $I=4$ and $6$; see Van~Isacker and
Heinze~\cite{IsHe14} and references therein.

\section{\label{sc:str}Stretch-scheme ground state}

The stretch-scheme ground state is
\begin{equation}\label{eq:sig}
  |\sigma\ket=|\{[(j_1j_5)\jdb(j_2j_6)\jdb]\jd
    [(j_3j_7)\jdb(j_4j_8)\jdb]\jd\}0\ket
\end{equation}
with $\jdb=2j$ and $\jd=4j-2$. Its image by the projection $P$ onto
the span of the states~\eqref{eq:ba} has the components
\begin{multline}\label{eq:sigc}  
  \bra\alpha|\sigma\ket
  =\bra\beta|\jeb\jeb\ket_{\je}\bra\gamma|\jeb\jeb\ket_{\je}\\
    \bra\je\je|\jd\jd\ket_{\jeb\jeb\jeb\jeb0} 
    \bra\jeb\jeb|\jdb\jdb\ket_{jjjj\jd}^2
\end{multline}
with $\jeb=2j-1$. The squared norm $\lVert P|\sigma\ket\rVert^2$ is
1.5\% for both $j$. This is much less than for $n=4$~\cite{Ne13},
where the corresponding squared norm is about 50\%. It means that
98.5\% of $|\sigma\ket$ carries irreps of the quasineutron $\times$
quasiproton $S(4)\times S(4)$ other than $[1^4]\times[1^4]$. Several
factors reduce $\lVert P|\sigma\ket\rVert^2$. First the unitary
nine-$j$ symbol $\bra\jeb\jeb|\jdb\jdb\ket_{jjjj\jd}$ in
Eq.~\eqref{eq:sigc} is about 0.7 for both $j$ and enters the squared
norm to the power of 4. Second $\bra\beta|\jeb\jeb\ket_{\je}$ vanishes
for odd $\je$ because a totally antisymmetric $|\je\beta\ket_n$ is
symmetric in $j_{12}$ and $j_{34}$. This gives another factor of about
$(1/2)^2$. Third the totally antisymmetric part of
$|\jeb\jeb\ket_{\je}$ for even $\je$ is typically about 1/3. With the
product $\bra\beta|\jeb\jeb\ket_{\je}\bra\gamma|\jeb\jeb\ket_{\je}$ in
Eq.~\eqref{eq:sigc} this factor enters $\lVert P|\sigma\ket\rVert^2$
to the power of 2.

Following Qi\etal~\cite{Qi11} I consider the state $\chi$ obtained by
normalization of $P|\sigma\ket$, that is,
\begin{equation}\label{eq:chi}
  |\chi\ket=\frac{P|\sigma\ket}{\lVert P|\sigma\ket\rVert}\,.
\end{equation}
Quite generally antisymmetrization in the quasiprotons and in the
quasineutrons of a product of $T=0$ states gives a $T=0$ state because
each factor in the product has a symmetry $[2^q]$ and the only $S(n)$
irrep with a Young frame with at most two columns containing a product
of such $S(2q)$ irreps is $[2^{n/2}]$. In particular because each pair
of quasinucleons with indices $i$ and $i+4$ in the
state~\eqref{eq:sig} has $T=0$ (symmetry $[2]$) the state $\chi$ has
$T=0$.

The $\Sp(2j+1)$ decomposition of $\chi$ is shown in Table~\ref{tb:tb}.
It is seen to be for both $j$ markedly different from those of the
calculated states $\psi$. In particular the irreps other than
$(s,t)=(0,0)$ and $(4,0)$, which are almost absent from $\psi$,
contribute 20\% and 41\%, respectively, of $\chi$, and $(s,t)=(6,1)$,
which makes only very small contributions to $\psi$---for $j=7/2$
vanishing within the numeric accuracy---gives for both $j$ the largest
of these contributions to $\chi$, about 20\% of the total. While the
calculated states are distributed in an approximate ratio 3\,:\,1 on
$(s,t)=(0,0)$ and $(4,0)$, these irreps have more equal weights in
$\chi$ with the contribution of $(s,t)=(4,0)$ being for $j=9/2$ even
the larger of the two. With 49\% and 25\%, respectively, for the two
$j$, the overlap of $\chi$ with the $s=0$ state is considerably less
for $n=8$ than for $n=4$, where it amounts to 62\% and 52\%,
respectively, as mentioned in the Introduction.

The overlaps of $\psi$ with $\chi$ are in the $1f_{7/2}$ shell about
the same as their overlaps with the $s=0$ state, 78\% and~77\% on
average over the interactions. In the $1g_{9/2}$ shell they are 60\%
on average over the interactions and the $s=0$ state is unambiguously
a better approximation to $\psi$ than $\chi$. The result
$|\bra\chi|\psi\ket|^2=66\%$ for the interaction QLW agrees with
Ref.~\cite{Qi11}.

When the subspace of the $I=T=0$ space belonging to a given
$\Sp(2j+1)$ irrep has a dimension larger than one, one may ask whether
the images of $\chi$ and $\psi$ by projection onto this subspace have
the same directions. This question is addressed in the last three
columns in Table~\ref{tb:tb}. Due to the numeric vanishing, mentioned
in the last paragraph of Sec.~\ref{sc:dec}, of the $(s,t)=(6,1)$
components of $\psi$ for $j=7/2$, this case is omitted. It is seen
that the directions are very much the same for $(s,t)=(4,0)$ and
almost as much so for $(s,t)=(8,0)$ in the $1g_{9/2}$ shell while the
situation is more ambiguous for $(s,t)=(6,1)$ in the $1g_{9/2}$ shell
with almost exactly orthogonal projected states for the interaction
ZE~I. Once more a similarity with the case $n=4$ studied in
Ref.~\cite{Ne13} is revealed: There, as well, the $(s,t)=(4,0)$
components of $\psi$ and $\chi$ have almost exactly the same
directions.

\section{\label{sc:oth}Product of stretch-scheme ground states}

In a $1f_{7/2}$ shell-model calculation for $^{48}$Cr with an
interaction interpolated from ZR~I and~II, Van Isacker~\cite{Is13}
finds that the state
\begin{equation}\label{eq:chi'}
  |\chi'\ket=\frac{P|\sigma'\ket}{\lVert P|\sigma'\ket\rVert}\,,
\end{equation}
with
\begin{equation}\label{eq:sig'}
  |\sigma'\ket=|[(j_1j_5)\jdb(j_2j_6)\jdb]0\ket
    \!\times\!|[(j_3j_7)\jdb(j_4j_8)\jdb]0\ket\,,
\end{equation}
makes up 92.7\% of the calculated ground state. The factors in the
product~\eqref{eq:sig'} are recognized as stretch-scheme ground states
of the $n=4$ system considered in Ref.~\cite{Ne13}. Like $\chi$ the
state $\chi'$ has $T=0$. The components of $P|\sigma'\ket$ in the
basis~\eqref{eq:ba} are
\begin{multline}\label{eq:sig'c}  
  \bra\alpha|\sigma'\ket
  =\sum_{j_aj_b }\bra\beta|j_aj_b \ket_{je}
    \bra\gamma|j_aj_b \ket_{\je}\\
    \bra\je\je|00\ket_{j_aj_b j_aj_b 0}
    \bra j_aj_a|\jdb\jdb\ket_{jjjj0}
    \bra j_b j_b |\jdb\jdb\ket_{jjjj0}\,,
\end{multline}
which gives $\Vert P|\sigma'\ket\Vert^2=0.9\%$ and 1.0\% for $j=7/2$
and~9/2, respectively. Properties of $\chi'$ are displayed in
Table~\ref{tb:tb}. The overlaps $|\bra\chi'|\psi\ket|^2$ are seen to
be considerably larger than $|\bra\chi|\psi\ket|^2$, about 93\% and
85\% in the $1f_{7/2}$ and $1g_{9/2}$ shells, respectively. The
overlaps calculated with the interactions ZR~I and~II are consistent
with Van Isacker's~\cite{Is13} with the interpolated interaction. The
overlap of $\chi'$ with the $s=0$ state is also closer to the one
found in Ref.~\cite{Ne13} for the $n=4$ antisymmetrized and normalized
stretch-scheme ground state. Like $\chi$ the images of $\chi'$ by
projection onto the $(s,t)=(4,0)$ and $(8,0)$ spaces have practically
the same directions as those of $\psi$. Its $\Sp(2j+1)$ decomposition
deviates, however, significantly from that of $\psi$, especially in
the $1g_{9/2}$ shell, although not quite as much as that of~$\chi$. In
particular $\chi'$ has like $\chi$ fairly large components of the
irreps other than $(s,t)=(0,0)$ and $(4,0)$, which are almost absent
in $\psi$.

The relative success of $\chi'$ in reproducing $\psi$ might be
understood from Qi's~\cite{Qi12a} observation that the $^{92}$Pd
ground state is well described in the $1g_{9/2}$ shell model as an
antisymmetrized and normalized product of two $^{96}$Gd ground states.
Because the $^{96}$Gd ground states have in this model a very large
overlap with the corresponding antisymmetrized and normalized
stretch-scheme ground state~\cite{Qi11,Ne13,ZeIs11,Is13}, one would
then anticipate that an antisymmetrized and normalized product of
copies of the latter has also a large overlap with the calculated
$^{92}$Pd ground states. That $\chi'$ is a better approximation to
$\psi$ in the $1f_{7/2}$ than in the $1g_{9/2}$ shell is in this
understanding consistent with the finding in Ref.~\cite{Ne13} that the
same holds for $n=4$ in the comparison of the calculated states and
the antisymmetrized and normalized stretch-scheme ground state. The
$n=8$ overlaps are indeed fairly close to the squares of the $n=4$
overlaps.

\section{\label{sc:sum}Summary}

In the $1f_{7/2}$ or $1g_{9/2}$ shell model with effective interaction
from the literature, I calculated the ground states of the system of
four neutrons and four protons or four neutron holes and four proton
holes, briefly four quasineutrons and four equivalent quasiprotons.
This is the single-$j$-shell configuration of the nuclei $^{48}$Cr,
$^{88}$Ru, and $^{92}$Pd. The calculated states $\psi$ were decomposed
into the irreps of the symplectic group $\Sp(2j+1)$, which are
characterized by the seniority $s$ and the reduced isospin $t$. Here
$j$ is the single-nucleon angular momentum, equal in the shells
considered to $7/2$ and $9/2$, respectively. The states $\psi$ are
found to be composed roughly of 75\% of $(s,t)=(0,0)$ and 25\% of
$(s,t)=(4,0)$ independently of the shell and the interaction. This is
similar to the case of two quasineutrons and two equivalent
quasiprotons studied in Ref.~\cite{Ne13}, where the corresponding
parts are about 80\% and 20\%. Other $\Sp(2j+1)$ irreps, which may
occur for $n=8$, make only very small contributions. This was
understood from the exact vanishing of the matrix elements of any
two-quasinucleon interaction between irreps with a difference in $s$
larger than 4 and small matrix elements between $(s,t)=(0,0)$ and
$(4,2)$ in the $1g_{9/2}$ shell. For $j=7/2$ also all matrix elements
involving $(s,t)=(6,1)$, and therefore these ground state components,
vanish within the numeric accuracy.

The ground state in the stretch scheme of Danos and
Gillet~\cite{DaGi67} was antisymmetrized in the quasineutrons and in
the quasiprotons. The antisymmetrized state is found to make up 1.5\%
of the total for both $j$'s. Following Qi\etal~\cite{Qi11}, I
considered the state $\chi$ given by normalization of this
antisymmetrized state. It is found to contain 20\% and 41\% of
$\Sp(2j+1)$ irreps other than $(s,t)=(0,0)$ and $(4,0)$ for $j=7/2$
and $9/2$, respectively, much unlike $\psi$. For both $j$'s the major
part of this contribution, about 20\% of the total in both cases,
resides in $(s,t)=(6,1)$. Unlike $\psi$ the irreps $(s,t)=(0,0)$ and
$(4,0)$ contribute roughly equally to $\chi$ and $(s,t)=(4,0)$ makes
for $j=9/2$, the larger of these two contributions.

The overlaps of $\psi$ with $\chi$ are found to be for $^{48}$Cr
similar to their overlaps with the $s=0$ state. For $^{88}$Ru and
$^{92}$Pd they are significantly less, so that the $s=0$ state is
there unambiguously a better approximation to $\psi$ than $\chi$. For
$^{92}$Pd and the interaction employed by Qi\etal\ in
Ref.~\cite{Qi11}, their result for $|\bra\chi|\psi\ket|^2$ is
confirmed.

The $\Sp(2j+1)$ irreps $(s,t)=(4,0)$ and $(6,1)$ have for both $j$
intersections of dimensions larger than one with the space with
angular momentum and isospin zero. So does the irrep $(s,t)=(8,0)$ for
$j=9/2$. The images of $\psi$ and $\chi$ by projection onto these
multidimensional spaces are found to have in a good approximation the
same directions for $(s,t)=(4,0)$ and $(8,0)$, whereas for $j=9/2$ and
$(s,t)=(6,1)$ the result in this respect varies with the interaction.
As to $(s,t)=(4,0)$, this is similar to the case of two quasineutrons
and two equivalent quasiprotons studied in Ref.~\cite{Ne13}. Due to
the aforesaid vanishing for $j=7/2$ of the $(s,t)=(6,1)$ components of
$\psi$, no comparison of directions is possible in this case.

The state $\chi'$ obtained by antisymmetrization and normalization of
the product of two stretch-scheme ground states of the system of two
quasineutrons and two equivalent quasiprotons was discussed briefly.
It has much larger overlaps with $\psi$ than $\chi$ but a deviating
$\Sp(2j+1)$ decomposition. The large overlaps $|\bra\chi'|\psi\ket|^2$
might be understood from Qi's observation~\cite{Qi12a} that the
$^{92}$Pd ground state is well described in the $1g_{9/2}$ shell model
as an antisymmetrized and normalized product of two $^{96}$Gd ground
states.

\bibliography{sen}

\end{document}